# Cross-Color Channel Perceptually Adaptive Quantization for HEVC

Lee Prangnell, Miguel Hernández-Cabronero and Victor Sanchez

*University of Warwick*
*Coventry, England, UK*
*{l.j.prangnell, m.hernandez-cabronero, v.f.sanchez-silva}@warwick.ac.uk*

**Abstract**: HEVC includes a Coding Unit (CU) level luminance-based perceptual quantization technique known as *AdaptiveQP*. AdaptiveQP perceptually adjusts the Quantization Parameter (QP) at the CU level based on the spatial activity of the pixel data in a luma Coding Block (CB). In this paper, we propose a novel cross-color channel adaptive quantization scheme which perceptually adjusts the CU level QP according to the spatial activity of pixel data in the constituent luma and chroma CBs; i.e., the combined spatial activity across all three color channels (the Y, Cb and Cr channels). Our technique is evaluated in HM 16 with 4:4:4, 4:2:2 and 4:2:0 YCbCr JCT-VC test sequences. Both subjective and objective visual quality evaluations are undertaken during which we compare our method with AdaptiveQP. Our technique achieves considerable coding efficiency improvements, with maximum BD-Rate reductions of 15.9% (Y), 13.1% (Cr) and 16.1% (Cb) in addition to a maximum decoding time reduction of 11.0%.

## 1. Introduction

Numerous psychophysical experiments reveal that the Human Visual System (HVS) is typically less sensitive to quantization-related distortions within regions of luminance and chrominance image data that comprise significant spatial variations [1]-[4]. Consequently, and in the context of video coding, higher levels of quantization can be applied in high spatial activity regions of frames in a sequence. This subsequently gives rise to useful bitrate reductions without incurring a perceptually discernable loss of reconstruction quality, which constitutes perceptual quantization.

AdaptiveQP, a $2N \times 2N$ CU level perceptual quantization technique in HEVC [5, 6], exploits this fact by applying a higher QP — relative to the slice level QP — to regions in a CU in which the luma CB consists of high spatial activity (pixel values); this typically results in coding efficiency improvements compared with Uniform Reconstruction Quantization (URQ) [5]-[7]. Conversely, a lower QP is employed in low spatial activity regions [5]. This lower CU QP selection, employed according to low spatial activity computations in a luma CB, can yield improved reconstruction quality compared with URQ [7]. AdaptiveQP achieves its objective by increasing or decreasing the QP of an entire $2N \times 2N$ CU based on the spatial activity of pixel data in a luma CB (without taking into account the data in the chroma CBs) [5, 6]. The fact that AdaptiveQP disregards the data in chroma Cb and Cr CBs during the CU QP selection process constitutes a significant shortcoming of this method. Our adaptive perceptual quantization method (*C-BAQ*) overcomes this shortcoming by accounting for both luma and chroma data in a $2N \times 2N$ CU.

CU level perceptually adaptive quantization, based on cross-color channel dependency for QP selection, has not been previously explored in HEVC research. However, perceptual quantization methods, similar to AdaptiveQP, have been previously proposed. The method in [8] exploits the luminance masking phenomenon of the HVS and applies it to HEVC. This technique is modeled on a Just Noticeable Distortion (JND) approach; it perceptually adjusts the QP based on JND and the average intensity of samples in luma CBs. The technique in [9] is also a JND perceptually adaptive quantization scheme that exploits luminance masking. This method is tailored for High Dynamic Range (HDR) input video signals.

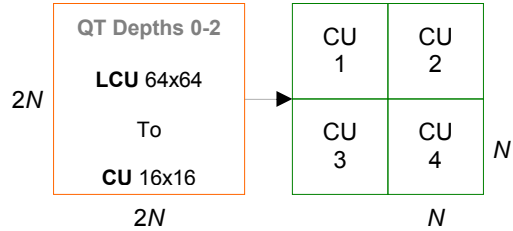

**Figure 1:** The CU size for which the QP is modified is 2*N*×2*N*. Both AdaptiveQP and C-BAQ operate at QT depth levels 0-2. When the split flag is enabled in HM, the 2*N*×2*N* CUs at QT depth levels 0-2 are partitioned into four constituent *N*×*N* CUs, where *N*=32 (level 0), *N*=16 (level 1) or *N*=8 (level 2). Note that CUs are always size 2*N*×2*N* or *N*×*N*. In other words and in contrast to CBs, CUs do not change in size due to chroma Cb and chroma Cr subsampling.

The proposed C-BAQ technique is a relatively simple, yet significant, improvement upon AdaptiveQP. In contrast to AdaptiveQP, C-BAQ accounts for the spatial activity of pixel data in both luma CBs and chroma CBs. Due to the fact that C-BAQ quantifies the population variance of the pixel intensities in the luma and chroma CBs, the 2*N*×2*N* CU level QP is perceptually adjusted according to the variances in all three CBs. Like AdaptiveQP, C-BAQ increases, or decreases, the CU level QP according to the population variances of the pixel data in the CU. Because of the cross-color channel dependency for QP selection in C-BAQ, the proposed technique has the potential to considerably improve coding efficiency without affecting the perceived visual quality in the reconstructed sequence. Furthermore, due to the potential decrease in the CU level QP for low spatial activity regions in a 2*N*×2*N* CU, overall reconstruction quality improvements, as quantified by PSNR increases, may be attained in cases where the luma and chroma CBs contain regions of data in which the population variances are low.

The rest of this paper is organized as follows. Section 2 includes technical information on the AdaptiveQP method. Section 3 provides technical details of the proposed C-BAQ technique. Section 4 includes the evaluations and results, in which C-BAQ is compared with AdaptiveQP. Section 5 provides a discussion of the evaluation. Finally, Section 6 concludes this paper.

## 2. AdaptiveQP in HEVC

Firstly, it is appropriate to distinguish the 2*N*×2*N* CU, the *N*×*N* CU and the CB. Assuming that the split flag is enabled in the HEVC HM reference software, the 2*N*×2*N* CU comprises four constituent *N*×*N* CUs (see Figure 1). The Largest Coding Unit (LCU) supports 64×64 samples and the Smallest Coding Unit (SCU) supports 8×8 samples. LCUs operate at QuadTree (QT) Depth Level=0 and SCUs operate at QT Depth Level=3 [11]-[13]. AdaptiveQP does not operate below QT Depth Level=2. The CU, at all QT depth levels, comprises three CBs (assuming that the input video data is not monochrome): one Y CB, one Cb CB and one Cr CB.

As previously mentioned in Section 1, AdaptiveQP is a luminance-based CU level perceptual quantization technique. It perceptually adjusts the QP of a 2*N*×2*N* CU based on the spatial activity of the pixel data in the four constituent *N*×*N* sub-blocks of a luma CB. More specifically, it quantifies the spatial activity based on the population variance of the pixel intensities in the sub-blocks of a luma CB. Therefore, a higher QP value is applied to luma CBs in which the variance is high (due to the aforementioned HVS masking effect). Conversely, a lower QP value is applied to luma CBs in which the variance is low. The CU level QP, denoted as *Q*, is computed in (1) [5]:

$$Q = q + \left[6 \times \log_2(n)\right] \tag{1}$$

where *q* corresponds to the slice level QP and where *n* refers to the normalized spatial activity of samples in a CB. Variable *n* is computed in (2):

$$n = \frac{f \cdot l + t}{l + f \cdot t} \tag{2}$$

where *f* is a scaling factor associated with the URQ QP adaptation range (denoted as *a*) regardless of the YCbCr color channel; note that *a* = 6 is the default value in JCT-VC HEVC HM. Variable *l* corresponds to the spatial activity of pixel values in a luma CB and variable *t* refers to the mean spatial activity for all 2*N*×2*N* CUs. Variables *f* and *l* are computed in (3) and (4), respectively:

$$f = 2^{\frac{a}{6}} \tag{3}$$

$$l = 1 + \min\left(\sigma^2_{Y,k}\right), \quad \text{where } k = 1, \ldots, 4 \tag{4}$$

where $\sigma^2_{Y,k}$ denotes the spatial activity of pixels values in sub-block *k* (of size *N*×*N*) in a luma CB. Variable $\sigma^2_{Y,k}$ is quantified as the population variance of luma pixel values, which is computed in (5):

$$\sigma^2_{Y,k} = \frac{1}{z}\sum_{i=1}^{z}(w_i - \mu_Y)^2 \tag{5}$$

where *z* denotes the number of pixel values in luma CB sub-block *k*. Variable $w_i$ corresponds to the *i*[th] sample in luma CB sub-block *k* and where $\mu_Y$ refers to the mean pixel intensity of luma CB sub-block *k*, which is computed in (6).

$$\mu_Y = \frac{1}{z}\sum_{i=1}^{z} w_i \tag{6}$$

## 3. Proposed C-BAQ Technique for HEVC

C-BAQ improves upon AdaptiveQP by accounting for the spatial activity of the pixel data in luma CBs, chroma Cb CBs and chroma Cr CBs. We achieve this by quantifying the population variance of luma and chroma pixel values in the corresponding CBs. The selection of the 2*N*×2*N* CU level QP is contingent upon the spatial activity of the data across all three CBs, which constitutes cross-color channel dependency for QP selection. C-BAQ perceptually adjusts the 2*N*×2*N* CU level QP according to the spatial activity of the data in each of the four constituent *N*×*N* sub-blocks of the luma and chroma CBs. Moreover, like AdaptiveQP, C-BAQ does not operate below QT Depth Level = 2 (see Figure 1). In terms of accounting for the pixel data in luma CBs, chroma Cb CBs and chroma Cr CBs, our technique is designed to derive a more appropriate QP selection for the 2*N*×2*N* CU as a whole. In essence and as previously implied, the primary objective of C-BAQ is to improve the perceptual quantization of the corresponding luma and chroma residual signals in order to considerably decrease overall bitrates without incurring a loss of perceptually discernible reconstruction quality. The CU level QP, denoted as $\tilde{Q}$, is computed in (7):

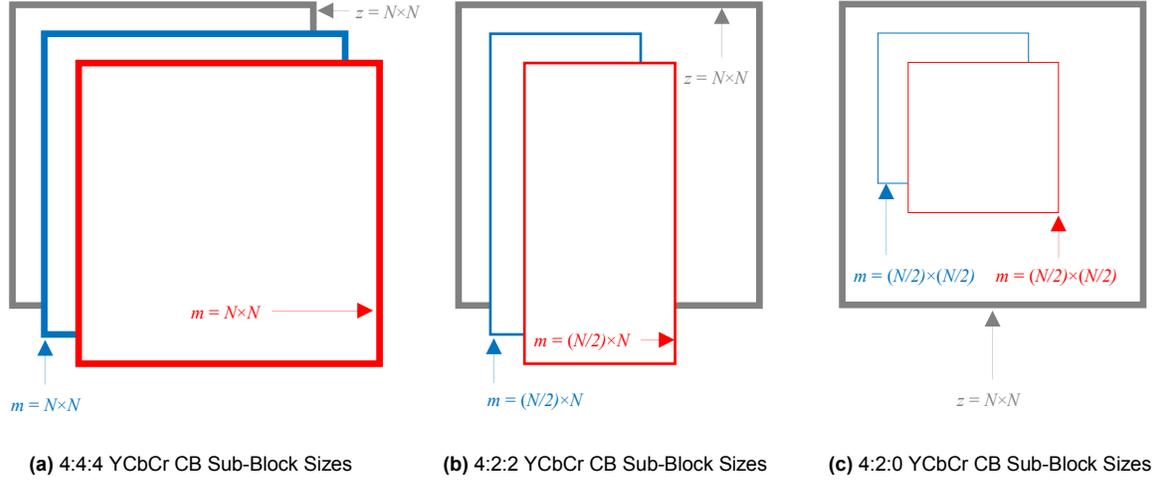

**Figure 2:** The sizes of sub-blocks in luma and chroma CBs in a 2N×2N CU in C-BAQ: Y (gray), Cb (blue), Cr (red). In C-BAQ, there are four constituent sub-blocks in the Y, Cb and Cr CBs in a 2N×2N CU. Each subfigure specifies the size of sub-blocks for different input video data: (a) for 4:4:4 YCbCr video data, the sub-block sizes for Y, Cb and Cr are all N×N, (b) for YCbCr 4:2:2 video data, the sub-block sizes are as follows: Y = N×N, Cb = (N/2)×N and Cr = (N/2)×N, (c) for YCbCr 4:2:0 video data, the sub-block sizes are as follows: Y = N×N, Cb = (N/2)×(N/2) and Cr = (N/2)×(N/2).

$$\tilde{Q} = q + \left\lceil 6 \times \log_2(\tilde{n}) \right\rceil \tag{7}$$

where $\tilde{n}$ denotes the normalized spatial activity of samples in both luma and chroma CBs. Variable $\tilde{n}$ is computed in (8):

$$\tilde{n} = \frac{f \cdot (l + b + d) + t}{(l + b + d) + f \cdot t} \tag{8}$$

where variables $b$ and $d$ correspond to the spatial activity of pixels in the chroma Cb and Cr CBs, respectively. Note that the HVS is typically more sensitive to gradations to the data in the luma channel. Moreover, the data in the chroma channels is susceptible to severe artifacts caused by very high levels of quantization. Therefore, in the HEVC standard the maximum QP permitted for chroma data is QP = 39 (chroma QP offset [14]) for YCbCr 4:2:0 chroma subsampled input video data [14]-[18]. Variables $b$ and $d$ are computed in (9) and (10), respectively.

$$b = 1 + \min(\sigma^2_{Cb,k}), \quad \text{where } k = 1,....,4 \tag{9}$$

$$d = 1 + \min(\sigma^2_{Cr,k}), \quad \text{where } k = 1,....,4 \tag{10}$$

where $\sigma^2_{Cb,k}$ and $\sigma^2_{Cr,k}$ refer to the spatial activity of pixels in sub-blocks $k$ in chroma Cb and Cr CBs, respectively. Variables $\sigma^2_{Cb,k}$ and $\sigma^2_{Cr,k}$ are computed as the population variance of Cb and Cr pixel values, respectively, given by (11) and (12), respectively (See Figure 2):

$$\sigma^2_{Cb,k} = \frac{1}{m} \sum_{i=1}^{m} (v_i - \mu_{Cb})^2 \tag{11}$$

$$\sigma^2_{Cr,k} = \frac{1}{m} \sum_{i=1}^{m} (j_i - \mu_{Cr})^2 \tag{12}$$

Table 1: BD-Rate results attained by the proposed C-BAQ technique compared with AdaptiveQP. The All Intra results are shown on the left and the Random Access results are shown on the right. Negative percentages indicate performance improvements of the proposed C-BAQ method in comparison with AdaptiveQP.

| C-BAQ versus AdaptiveQP (YCbCr 4:2:0) – All Intra | | | | C-BAQ versus AdaptiveQP (YCbCr 4:2:0) – Random Access | | | |
|---|---|---|---|---|---|---|---|
| Sequence | BD-Rate % | | | Sequence | BD-Rate % | | |
| | Y | Cb | Cr | | Y | Cb | Cr |
| FourPeople (8-bit) | −9.5 | −8.6 | −9.9 | FourPeople (8-bit) | −8.7 | −7.5 | −8.0 |
| KristenAndSara (8-bit) | −14.3 | −12.3 | −12.5 | KristenAndSara (8-bit) | −15.5 | −12.8 | −11.8 |
| ParkScene (8-bit) | −5.4 | −8.0 | −7.8 | ParkScene (8-bit) | −4.0 | −6.1 | −6.2 |
| Traffic (8-bit) | −8.6 | −10.6 | −13.5 | Traffic (8-bit) | −4.9 | −7.0 | −9.0 |
| C-BAQ versus AdaptiveQP (YCbCr 4:2:2) – All Intra | | | | C-BAQ versus AdaptiveQP (YCbCr 4:2:2) – Random Access | | | |
| Sequence | BD-Rate % | | | Sequence | BD-Rate % | | |
| | Y | Cb | Cr | | Y | Cb | Cr |
| PeopleOnStreet (8-bit) | −9.8 | −13.4 | −9.6 | PeopleOnStreet (8-bit) | −5.3 | −5.5 | −3.9 |
| DuckAndLegs (10-bit) | −6.0 | −4.2 | −8.3 | DuckAndLegs (10-bit) | −8.0 | −9.2 | −11.0 |
| ParkScene (10-bit) | −9.7 | −9.2 | −16.1 | ParkScene (10-bit) | −7.5 | −12.8 | −13.5 |
| Traffic (10-bit) | −9.2 | −12.2 | −15.3 | Traffic (10-bit) | −5.0 | −9.3 | −11.4 |
| C-BAQ versus AdaptiveQP (YCbCr 4:4:4) – All Intra | | | | C-BAQ versus AdaptiveQP (YCbCr 4:4:4) – Random Access | | | |
| Sequence | BD-Rate % | | | Sequence | BD-Rate % | | |
| | Y | Cb | Cr | | Y | Cb | Cr |
| PeopleOnStreet (8-bit) | −11.8 | −14.0 | −9.0 | PeopleOnStreet (8-bit) | −6.7 | −7.1 | −6.4 |
| DuckAndLegs (10-bit) | −14.0 | −7.0 | −11.2 | DuckAndLegs (10-bit) | −15.9 | −13.1 | −16.1 |
| ParkScene (10-bit) | −15.6 | −8.7 | −19.3 | ParkScene (10-bit) | −12.0 | −16.4 | −17.0 |
| Traffic (10-bit) | −11.1 | −13.4 | −15.9 | Traffic (10-bit) | −5.6 | −11.3 | −11.9 |

where $m$ refers to the number of pixels in sub-blocks $k$ in Cb and Cr CBs (see Figure 2). Variables $v_i$ and $j_i$ correspond to the $i^{th}$ samples in Cb CB sub-block $k$ and Cr CB sub-block $k$, respectively. Variables $\mu_{Cb}$ and $\mu_{Cr}$ denote the mean pixel values in Cb CB sub-block $k$ and Cr CB sub-block $k$, respectively, which are quantified in (13) and (14), respectively.

$$\mu_{Cb} = \frac{1}{m}\sum_{i=1}^{m} v_i \tag{13}$$

$$\mu_{Cr} = \frac{1}{m}\sum_{i=1}^{m} j_i \tag{14}$$

## 4. Experimental Evaluations

We evaluate C-BAQ and compare it with AdaptiveQP. We integrate C-BAQ into HEVC HM 16.7 and undertake thorough evaluations that correspond, as closely as possible, to JCT-VC's Common Test Conditions and Software Reference Configurations [19]. C-BAQ is a HVS-based perceptual quantization technique; therefore, it is of considerable importance to undertake a subjective visual quality evaluation in addition to an objective visual quality evaluation. Note that the subjective evaluation is vital because it allows us to fairly assess the reconstruction quality of the C-BAQ coded sequences versus the AdaptiveQP coded sequences; the JCT-VC test sequences, and the corresponding bit depths, are shown in Table 1. The experimental setup, which applies to both C-BAQ and AdaptiveQP, is summarized in the following list:

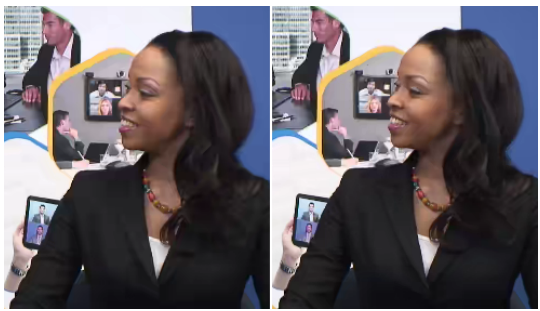 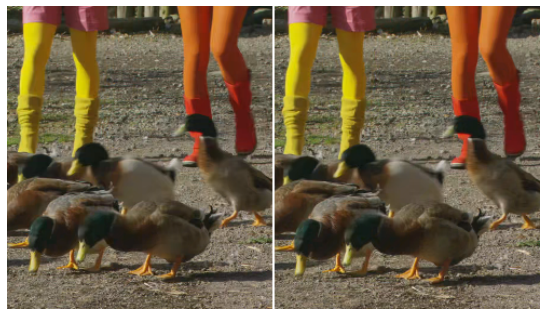

**Figure 3:** Comparison of an intra-predicted frame from the *KristenAndSara* 4:2:0 sequence at QP 37 (All Intra): (a) coded with C-BAQ — 5580.99 Kbps, (b) coded with AdaptiveQP — 5998.07 Kbps.

**Figure 4:** Comparison of an inter-predicted frame from the *DuckAndLegs* 4:4:4 sequence at QP 37 (Random Access): (a) coded with C-BAQ — 2390.49 Kbps, (b) coded with AdaptiveQP — 2591.75 Kbps.

- **Evaluation Metrics:** Subjective Visual Quality Evaluation and BD-Rate [20].
- **QPs:** 22, 27, 32 and 37 (Objective Evaluation) [19]. **QP:** 37 (Subjective Evaluation).
- **Encoding Configurations:** All Intra and Random Access.
- **Encoding Profiles:** Main, Main_422_10, Main_444_10, Main_444, Main_422_10_Intra, Main_444_10_Intra and Main_444_Intra.

Five experienced researchers in the field video coding performed a series of thorough subjective visual quality evaluations — i.e., C-BAQ coded videos versus AdaptiveQP coded videos. The participants analyzed the visual differences in the reconstructed sequences in side-by-side comparisons for all JCT-VC sequences coded (as shown in Table 1). Note that the participants were shown reconstructed sequences coded using QP 37 only. The reason for this is to establish if the participants were able to discern visual differences between C-BAQ coded sequences and AdaptiveQP coded sequences when the quantization-induced compression artifacts should be most visible (i.e., with the highest QP value used). 81.25% of the participants perceived either no visual quality differences between C-BAQ coded sequences and AdaptiveQP coded sequences, or C-BAQ coded sequences were perceived to be superior (see Figure 3 and Figure 4). The remaining 18.75% of participants viewed the AdaptiveQP coded sequences to be superior.

In both the All Intra and Random Access objective visual quality evaluations, considerable coding efficiency improvements are attained by C-BAQ in comparison with AdaptiveQP, as measured by BD-Rate reductions (see Table 1) [20]. BD-Rate percentages quantify bitrate measurements (e.g., decreases, no changes or increases in bitrate) when the reconstruction quality, as measured by the Peak Signal to Noise Ratio (PSNR) metric, is the same in both techniques tested [20]. High coding efficiency improvements are accomplished by C-BAQ in the All Intra evaluations, which are as follows. In the YCbCr 4:2:0 tests, BD-Rate reductions of 14.3% (Y), 12.3% (Cb) and 12.5% (Cr) are achieved on the 8-bit KristenAndSara sequence (Main encoding profile). In the YCbCr 4:2:2 tests, BD-Rate reductions of 9.2% (Y), 12.2% (Cb) and 15.3% (Cr) are attained on the 10-bit Traffic sequence (Main_422_10_Intra encoding profile). In the YCbCr 4:4:4 tests, noteworthy BD-Rate reductions of 15.6% (Y), 8.7% (Cb) and 19.3% (Cr) are accomplished on the 10-bit ParkScene sequence (Main_444_10_Intra encoding profile). Likewise, significant coding efficiency improvements are achieved by C-BAQ in the Random Access evaluations, which are as follows. BD-Rate reductions of 15.5% (Y), 12.8% (Cb) and 11.8% (Cr) are accomplished on the YCbCr 4:2:0 8-bit KristenAndSara sequence (Main encoding profile). BD-Rate reductions of 7.5% (Y), 12.8% (Cb) and 13.5% (Cr) are achieved on the YCbCr 4:2:2 10-bit ParkScene sequence (Main_422_10 encoding profile). Finally, in the YCbCr 4:4:4 simulations, considerable BD-Rate reductions of 15.9% (Y), 13.1% (Cb) and 16.1% (Cr) are attained on the 10-bit DuckAndLegs sequence (Main_444_10 encoding profile).

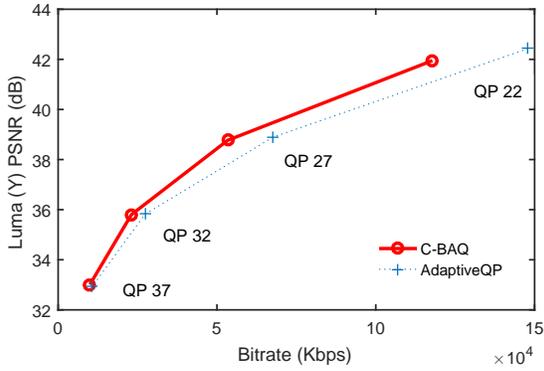

**Figure 5:** C-BAQ bitrate reductions (luma channel) in comparison with AdaptiveQP on the 4:4:4 10-bit test sequence *ParkScene* (using the Main_444_10_Intra RExt profile including the All Intra configurations).

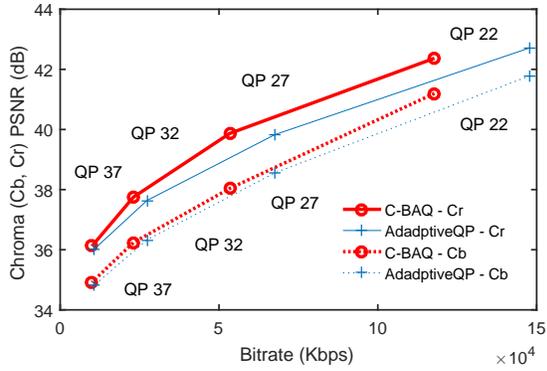

**Figure 6:** C-BAQ bitrate reductions (chroma channels) in comparison with AdaptiveQP on the 4:4:4 10-bit test sequence *ParkScene* (using the Main_444_10_Intra RExt profile including the All Intra encoding configurations).

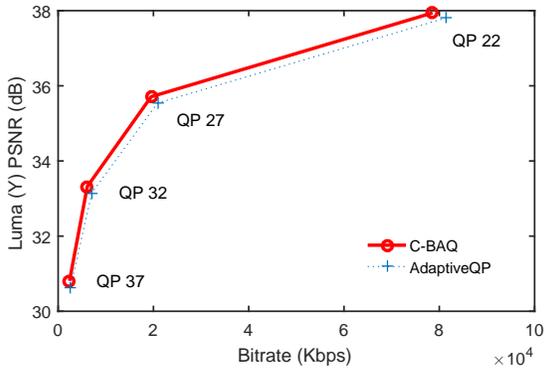

**Figure 7:** C-BAQ bitrate reductions (luma channel) in comparison with AdaptiveQP on the 4:4:4 10-bit test sequence *DuckAndLegs* (using the Main_444_10 RExt profile including the Random Access configurations).

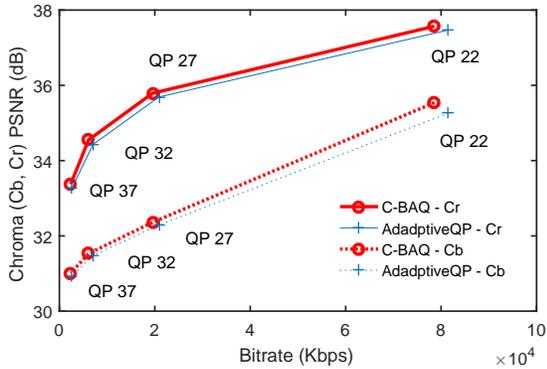

**Figure 8:** C-BAQ bitrate reductions (chroma channels) in comparison with AdaptiveQP on the 4:4:4 10-bit test sequence *DuckAndLegs* (using the Main_444_10 RExt profile including the Random Access configurations).

Figure 5 and Figure 6 include plots which illustrate the coding efficiency improvements achieved by C-BAQ in the YCbCr 4:4:4 10-bit ParkScene sequence test. Moreover, Figure 7 and Figure 8 include plots which highlight the bitrate reductions obtained by C-BAQ in the YCbCr 4:4:4 10-bit DuckAndLegs sequence tests.

In the majority of cases, the encoding time and decoding time performances of C-BAQ proved to be consistently superior in comparison with those of AdaptiveQP. In the All Intra tests, the most significant encoding time improvement is achieved in the YCbCr 4:4:4 PeopleOnStreet sequence test, with a reduction of 6.0%. The most noteworthy decoding time reduction is achieved in the YCbCr 4:2:2 ParkScene sequence test, with a reduction of 11.0%. In the Random Access tests, the overall differences in encoding times and decoding times between C-BAQ and AdaptiveQP proved to be marginal. The most significant encoding time improvement achieved by C-BAQ is in the YCbCr 4:2:0 KristenAndSara sequence test, with a reduction of 0.9%. A moderate decoding time is attained in the YCbCr 4:2:2 ParkScene sequence test, with a reduction of 3.4%.

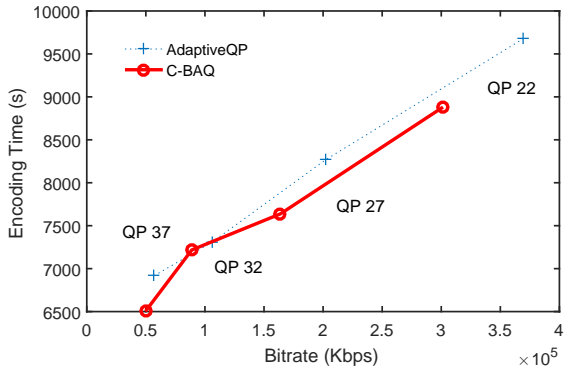
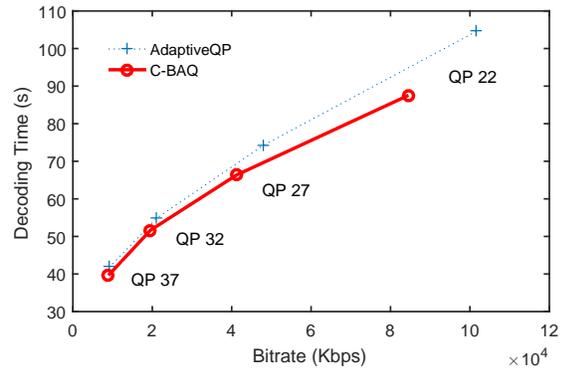

**Figure 9:** Encoding time improvements attained by the proposed C-BAQ technique compared with AdaptiveQP on the 4:4:4 8-bit test sequence *PeopleOnStreet* (using the Main_444_Intra RExt profile including the All Intra configurations).

**Figure 10:** Decoding time improvements attained by the proposed C-BAQ technique compared with AdaptiveQP on the 4:2:2 10-bit test sequence *ParkScene* (using the Main_422_10_Intra RExt profile including the All Intra configurations).

## 5. Discussion

In the vast majority of cases (81.25%), the participants in the subjective evaluation either confirmed that they noticed no visual quality differences between C-BAQ coded sequences versus AdaptiveQP sequences, or they perceived C-BAQ coded sequences to be superior — even though C-BAQ coded sequences are considerably lower in terms of bitrate. Therefore, in this discussion, we focus on the objective visual quality evaluation results.

As shown in Table 1, C-BAQ achieves superior coding efficiency when applied to versions of a test sequence in which each CU contains a greater degree of chroma data. In other words, C-BAQ is most effective when applied to the YCbCr 4:4:4 version of a sequence. It is less effective when applied to the YCbCr 4:2:2 version and, moreover, less effective still when applied to the YCbCr 4:2:0 version; we observed this behavior in both the All Intra and Random Access tests. For example, the coding efficiency performance of C-BAQ on the 4:4:4 versions of the ParkScene and Traffic sequences outperform the results obtained for the 4:2:2 and 4:2:0 versions of the same sequences. Furthermore, the bit depth of the YCbCr input video data is not a significant factor in terms of predicting whether or not C-BAQ will achieve a superior coding efficiency performance. Irrespective of the bit depth of the input video data, C-BAQ consistently achieves better results when applied to the 4:4:4 version of a sequence; it is palpably evident that C-BAQ yields inferior results on subsampled chroma versions of input video data. This is due to the disparity of sizes between the luma CB and the smaller chroma CBs in the 4:2:2 and 4:2:0 versions of a sequence (see Figure 2). This disparity affects the CB sample variance computations and, thus, results in the selection of a less appropriate cross-color channel CU level QP. The KristenAndSara and FourPeople test sequences exist in the 4:2:0 format only (see Table 1); therefore, we were unable to undertake evaluations on the 4:4:4 and 4:2:2 versions of these sequences.

In the All Intra tests, C-BAQ obtains significant encoding time performance improvements in addition to considerable decoding time reductions (see Figure 9 and Figure 10). Moderate encoding time and decoding time reductions are achieved in the Random Access tests. When high levels of spatial activity are detected in luma and chroma CBs, C-BAQ quantizes these areas with an increased QP — relative to AdaptiveQP's QP selection — without inducing a perceptible loss of visual quality in the reconstructed sequence; therefore, encoding times are subsequently reduced. Consequently, C-BAQ achieves decreased decoding times because fewer bits need to be decoded; this is because of the higher QP value selected during the encoding process.

In some cases, C-BAQ achieves reconstruction quality improvements in comparison with AdaptiveQP, as quantified by increases in PSNR values (see the plots in Figure 7 and Figure 8, for example). As stated in Section 1, both AdaptiveQP and C-BAQ can decrease the QP at the CU level, relative to the slice level QP, if the population variance of pixel data in any given CB is low. This may result in potential reconstruction quality improvements. We have established that AdaptiveQP takes into account pixel data in luma CBs only; this means that the QP of an entire CU is adjusted according to the variance of this data in the luma CB only. Therefore, the variances of the pixel data in the chroma CBs is completely disregarded, which leaves room for improvement (particularly for YCbCr 4:4:4 input video data). Due to the fact that C-BAQ takes into account both luma and chroma data, the cross-color channel dependency for QP selection equates to the fact that the variances in all three CBs in a CU are taken into account during CU level QP selection. If, for example, the variances in the Y Cb, the Cb CB and the Cr CB in any given CU are low, then the QP will be decreased, thus potentially resulting in reconstruction quality improvements.

## 6. Conclusions and Future Work

A novel CU level color-based perceptually adaptive quantization scheme, named *C-BAQ*, is proposed for the HEVC standard to potentially replace the AdaptiveQP technique. C-BAQ accounts for the spatial activity of Y, Cb and Cr pixel data, in the corresponding CBs, in order to select a more appropriate $2N \times 2N$ CU level QP during the coding process; this is achieved by employing a cross-color channel dependency for QP selection mechanism. C-BAQ was subsequently implemented into HEVC HM 16.7 for the purpose of undertaking experimental evaluations. Thorough visual quality evaluations — both subjective and objective — were undertaken, during which we compared C-BAQ with AdaptiveQP. In the subjective visual quality evaluation, 81.25% of the participants either discerned no differences between C-BAQ coded sequences and AdaptiveQP coded sequences, or C-BAQ coded sequences were interpreted to be superior. In the objective visual quality evaluation, and in comparison with AdaptiveQP, C-BAQ achieves outstanding coding efficiency improvements, with a maximum BD-Rate reduction of 15.9% (Y), 13.1% (Cr) and 16.1% (Cb). Improved encoding times and decoding times are also achieved, with maximum reductions of 6.0% and 11.0%, respectively.

In terms of future work related to C-BAQ, it is desirable to potentially develop a HVS-based JND model for C-BAQ that provides a more comprehensive and quantifiable account of the potential perceptual redundancies in the different color channels within the YCbCr color space. In contrast to the ISO/CIE CIELAB color space, for example, the YCbCr color space is not perceptually uniform, which equates to the fact that the Y, Cb and Cr channels cannot be considered as perceptually equal. The luma channel (Y) in the YCbCr color space is a gamma corrected weighted sum of tristimulus intensity values in the Red, Green, Blue (RGB) color space (the coefficients of which are presently standardized by ITU-R with Recommendation BT.2020). This means that the luma channel is an achromatic color channel that corresponds to the human perception of the brightness of color. Conversely, the chroma Cb and chroma Cr channels are color difference channels — blue difference and red difference, respectively — with reference to the luma channel; the chroma channels correspond to the human perception of the colorfulness of color. Therefore, it is important to take these factors into account when developing potential extended contributions to the proposed C-BAQ method. Moreover, potential extensions to this work should also include exhaustive objective quality evaluations and subjective quality evaluations.

# 7. References

[1] A. N. Netravali, N. J. Holmdel and B. Prasad, "Adaptive quantization of picture signals using spatial masking," *Proc. IEEE*, vol. 65, no. 4, pp. 536-548, 1977.

[2] D. J. Connor, R. C. Brainard, and J. O. Limb, "lntraframe Coding for Picture Transmission," *Proc. IEEE*, vol. 60, no. 7, pp. 779-791, 1972.

[3] A. N. Netravali and C. B. Rubinstein, "Quantization of color signals," *Proc. IEEE*, vol. 65, no. 8, pp. 1177-1187, 1977.

[4] S. W. Cheadle and S. Zeki, "Masking within and across visual dimensions: Psychophysical evidence for perceptual segregation of color and motion," *Visual Neuroscience*, vol. 28, no. 5, pp. 445-451, 2011.

[5] K. McCann, C. Rosewarne, B. Bross, M. Naccari, K. Sharman and G. J. Sullivan (Editors), "High Efficiency Video Coding (HEVC) Test Model 16 (HM 16) Encoder Description," in *JCT-VC R1002, 18th Meeting of JCT-VC*, Sapporo, JP, 2014, pp. 1-59.

[6] M. Naccari and M. Mrak, "Perceptually Optimized Video Compression," *Elsevier Academic Press Library in Signal Processing*, vol. 5, pp. 155-196, 2014.

[7] K. Sato, "On LBS and Quantization," in *JCT-VC D308, 4th Meeting of JCT-VC*, Daegu, KR, 2011, pp. 1-12.

[8] M. Naccari and M. Mrak, "Intensity Dependent Spatial Quantization with Application in HEVC," *IEEE Int. Conf. Multimedia and Expo*, San Jose, CA, 2013, pp. 1-6.

[9] Y. Zhang, M. Naccari, D. Agrafiotis, M. Mrak and D. Bull, "High Dynamic Range Video Compression Exploiting Luminance Masking," *IEEE Trans. Circuits Syst. Video Techn.*, vol. 26, no. 5, pp. 950-964, 2016.

[10] G. Sullivan, J-R. Ohm, W. Han and T. Wiegand, "Overview of the High Efficiency Video Coding (HEVC) Standard," *IEEE Trans. Circuits Syst. Video Technol.*, vol. 22, no. 12, pp. 1649-1668, 2012.

[11] V. Sze, M. Budagavi and G. J. Sullivan, "Block Structures and Parallelism Features in HEVC," in *High Efficiency Video Coding (HEVC): Algorithms and Architecture*, Springer, 2014, pp. 49-91.

[12] M. Wein, "Coding Structures," in *High Efficiency Video Coding: Coding Tools & Specification*, Springer, 2015, pp. 101-132.

[13] ITU-T Rec. H.265/HEVC (Version 3) | ISO/IEC 23008-2, Information Technology – Coding of Audio-Visual Objects," *ITU-T/ISO/IEC*, 2015.

[14] S. Liu and K. Sato, "Support of ChromaQPOffset in HEVC," in *JCT-VC G509, 7th Meeting of JCT-VC*, Geneva, CH, 2011, pp. 1-8.

[15] G. J. Sullivan, S. Kanumuri, J- Xu and Y. Wu, "Chroma QP range extension," in *JCT-VC J0342, 10th Meeting of JCT-VC*, Stockholm, SE, 2012, pp. 1-8.

[16] B. Bross, G. J. Sullivan, T. K. Tan and Y-K Wang, "Proposed Editorial Improvements for High efficiency video coding (HEVC)," in *JCT-VC K0383, 11th Meeting of JCT-VC*, Shanghai, CN, 2012, pp. 1-256.

[17] D. Flynn, M. Naccari, C. Rosewarne, K. Sharman, J. Sole, G. Sullivan and T. Suzuki, "High Efficiency Video Coding (HEVC) Range Extensions text specification: Draft 7," in *JCT-VC Q1005, 17th Meeting of JCT-VC*, Valencia, ES, 2014, pp. 1-330.

[18] M. Wein, "Video Coding Fundamentals," in *High Efficiency Video Coding: Coding Tools & Specification*, Springer International Publishing, 2015, pp. 23-72.

[19] F. Bossen, "Common test conditions and software reference configurations," in *JCT-VC L1100, 12th Meeting of JCT-VC*, Geneva, CH, 2013, pp. 1-4.

[20] G. Bjøntegaard, "Improvements of the BD-PSNR model," in *VCEG-AI11, 35th Meeting of VCEG*, Berlin, DE, 2008, pp. 1-8.